\documentclass[letterpaper,twocolumn,10pt]{article}
\pdfoutput=1
\usepackage{styles/usenix-2020-09}

\usepackage[normalem]{ulem}
\usepackage[ruled,vlined]{algorithm2e}
\usepackage{listings}
\usepackage{multirow}

\usepackage[sort&compress,numbers]{natbib}
\usepackage{amsmath,amssymb,amsfonts}
\usepackage{graphicx, subcaption}
\usepackage{textcomp}
\usepackage{xcolor}
\usepackage{balance}
\usepackage{soul}
\usepackage{textcomp}
\usepackage{tabularx}
\usepackage{algpseudocode}
\usepackage{xspace}
\usepackage{xparse}
\usepackage{courier}

\begin{document}
\date{}

\title{Compiler-Guided Throughput Scheduling for Many-core Machines}

\author{
{\rm Girish Mururu} \\
Google
\and
{\rm Sharjeel Khan} \\
Georgia Institute of Technology
\and
{\rm Bodhisatwa Chatterjee} \\
Georgia Institute of Technology
\and
{\rm Chao Chen} \\
Amazon
\and
{\rm Chris Porter} \\
Georgia Institute of Technology
\and
{\rm Ada Gavrilovska} \\
Georgia Institute of Technology
\and
{\rm Santosh Pande} \\
Georgia Institute of Technology
}

\maketitle

\thispagestyle{empty}

\begin{abstract}

Typical schedulers in multi-tenancy environments make use of reactive, feedback-oriented mechanisms based on performance counters to avoid resource contention but suffer from detection lag and loss of performance. 
In this paper, we address these limitations by exploring the utility of {\it predictive analysis} through dynamic forecasting of applications' resource-heavy regions during its execution. 
Our compiler framework classifies loops in programs and leverages traditional compiler analysis along with learning mechanisms to quantify their behaviour. Based on the predictability of their execution time, it then inserts different types of beacons at their entry/exit
points. 
The information produced by beacons in multiple processes is aggregated
and analyzed by the proactive scheduler to respond to
the anticipated workload requirements.
For throughput environments,
we develop a framework that
demonstrates high-quality predictions and
improvements in throughput over CFS by \textbf{76.78\%} on an average and up to \textbf{3.2x} on Amazon Graviton2 Machine
on consolidated workloads across 45 benchmarks.
\end{abstract}

\section{Introduction}
\label{sec:intro}
Modern systems offer a tremendous amount of computing resources by assembling a very large number of cores. However, when they are deployed in data-centers, it is very challenging to fully exploit such computing capability due to their shared memory resources. The challenge is that on one hand, there is a need to co-locate a large number of workloads to maximize the system utilization (\& thus minimizing the overall cost), and on the other hand, this workload co-location causes processes to interfere with each other by competing for shared cache resources. This problem is further exacerbated by the fact that modern machine learning \& graph analytics workloads exhibit \textit{diverse resource demands} throughout their execution. In particular, the resource requirements of processes vary not only across applications, but also within a single application across its various regions/phases as it executes \cite{fried2020caladan, iorgulescu2018perfiso}. 

Prior works on resource management \& utilization \cite{chen2019parties, ebrahimi2010fairness, paragon, bubbleflux, priyanka:socc14,javadi2019scavenger} do not account for the dynamic phase behaviour in modern workloads and use feedback-driven reactive approaches instead. These include approaches that rely on resource usage history or current resource contention (monitored using hardware performance counters) and then \textit{reacting} to contention by invoking suitable scheduling mechanisms. However, majority of modern workloads are input data dependent and do not exhibit highly regular or repetitive behavior across different input sets. Thus, neither history-based methods nor feedback directed methods provide sufficiently accurate predictions for such workloads. Furthermore, these reactive approaches detect a resource-intense execution phase only after it already occurred, and this leads to two more related drawbacks. First, by the time the phase is detected, it may be too late to act, especially if the detection lag is significant or phase duration is short (a phase might be end by the time it is detected). Second, as a result of this detection lag, the application state might have already bloated in terms of its resource consumption; the damage to other co-executing processes' cache state might be already done, and it might be prohibitively expensive to mitigate the offending process because of the large state and its cache affinity can be lost by migrating or pausing. In addition, these approaches cannot predict the duration of the phase, nor the sensitivity of its forthcoming resource usage, thus significantly limiting the chances of further optimizations to co-location of processes.

To efficiently co-locate processes and maximize system throughput, we propose \textbf{Beacons Framework} to predict the resource requirement of applications based on a combination of compiler analysis and machine learning techniques that enables pro-active scheduling decisions. This framework consists of \textbf{Beacons Compilation Component}, that leverages the compiler to analyze programs and predict the resource requirements, \& \textbf{Beacons Runtime Component}, which includes the pro-active \textit{Beacons Scheduler}, that uses these analysis to undertake superior scheduling decisions. The beacons scheduler relies on the compiler to insert ``beacons''(specialized markers) in the application at strategic program points to periodically produce and/or update details of anticipated (forthcoming) resource-heavy program region(s). The beacons framework classifies loops in programs based on cache usage and predictability of their execution time and inserts different types of beacons at their entry/exit points.  Moreover, the beacons scheduler augments imprecise information through the use of performance counters to throttle higher concurrency when needed. The novelty of this scheme is that through the use of \textit{loop timing}, \textit{memory footprint}, and \textit{reuse information} provided via compiler analysis combined with machine learning mechanisms, the scheduler is able to precisely yet aggressively maximize co-location or concurrency without incurring cache degradation.

We evaulated \textbf{Beacons Framework} on a Amazon Graviton2~\cite{graviton2} machine on consolidated benchmark suites from \textit{Machine Learning Workloads} \cite{huang2017densely,hinton2012improving,deng2009imagenet,he2016deep}, \textit{PolyBench}~\cite{yuki2015polybench}, a numerical computational benchmark, and \textit{Rodinia}~\cite{che2009rodinia}, a heterogeneous compute-based kernel based on diverse domains. Our experiments demonstrate that the end result of such a high throughput oriented framework is that it effectively schedules a large number of jobs and improves throughput over CFS (Completely Fair Scheduler, a production scheduler bundled with Linux distributions) by \textbf{76.78\%} on average (geometric mean) and up to \textbf{3.29x} (\S\ref{sec:eval}). Overall, the main contributions of our work can be summarized as follows:
\begin{itemize}
    \item We present a compiler-directed predictive analysis framework that determines resource requirements of dynamic modern workloads in order to aid dynamic resource management.
    \item We show that an application's resource requirement can be determined by its \textit{loop-timings} (\S\ref{bec:timing}) which in turn depend upon the prediction of expected loop iterations (\S\ref{bec:iter}), \textit{memory footprints} (\S\ref{sec:memfp}) and \textit{data-reuse behaviour} (\S\ref{bec:reuse}). We show how these attributes can be estimated by combining  compiler analysis with machine learning techniques in Beacons Compilation Component (\S\ref{beacons:compiler}).
    \item To predict precise loop timing for proactive scheduling decisions, we develop a detailed loop classification scheme that covers all possible types of loops in modern workloads. Based on this scheme, we develop a framework that analyzes such loops by capturing both the data-flow and control-flow aspects (\S\ref{bec:iter}).
    \item We design and implement a lightweight \textit{Beacons Runtime Component} (\S\ref{beacons_sch}), which includes the proactive throughput-oriented Beacons Scheduler (\S\ref{bec:sch}), that aggregates the information generated by multiple processes and allocate resources to the processes and enables communication between the application and the scheduler. 
\end{itemize}

\section{Motivation \& Overall Framework}
\label{sec:motiv}
Maximizing system throughput by efficient process scheduling in a multi-tenant execution setting (data centers, cloud, etc) for modern workloads is challenging. The reason for this is that these execution scenarios consists of thousand of jobs, typically in batches. In such cases, completion time of each individual job itself is of no value, but the completion time of either whole batch of jobs, or the number of jobs completed per unit time (if the jobs are an incoming stream) is of utmost priority \cite{amvrosiadis2018diversity}. 

\subsection{Motivating Example: Reactive Vs Proactive}
Consider an execution scenario in a multi-tenant enviroment, where we have the popular convolutional neural network (CNN) \textit{Alexnet}, being trained on different input sets with varied parameters (dropout rate, hidden units, etc) concurrently, as a batch. During the training, several other jobs (matrix multiplication computations on different matrices) co-execute, having access to the same shared resources (LLC, etc), but on different cores. Table \ref{tab:t} shows the comparison of execution time of this scenario with \textbf{(1)} widely used production Linux scheduler \textbf{CFS} \cite{kobus2009completely}, \textbf{(2)} a performance-counter based reactive scheduler \textbf{Merlin} \cite{priyanka:socc14} and \textbf{(3)} proposed \textbf{Beacons Framework} with Beacons Scheduler.      

\begin{table}[!ht]
\resizebox{\columnwidth}{!}{
\begin{tabular}{|l|l|l|l|l|}
\hline
\multirow{2}{*}{\textbf{\begin{tabular}[c]{@{}l@{}}Constituent Processes\\ executing as a Batch\end{tabular}}} & \multirow{2}{*}{\textbf{Total Processes}} & \multicolumn{3}{l|}{\textbf{Execution Time (sec)}} \\ \cline{3-5} 
 &  & \textbf{CFS} & \textbf{Merlin} & \textbf{Beacons} \\ \hline
Alexnet Training & 20 & \multirow{2}{*}{249.43} & \multirow{2}{*}{358.23} & \multirow{2}{*}{100.58} \\ \cline{1-2}
Matrix Multiplication (2mm) & 133768 &  &  &  \\ \hline
\end{tabular}}
\caption{Alexnet Training and 2mm (Polybench) executing as a batch. The 2mm process can be considered as a small process compared to Alexnet, and hence can act as hogging system resources with sheer numbers}
\label{tab:t}
\end{table}

As we can see from Table \ref{tab:t}, the schedulers like CFS that are agnostic to the diverse requirements and overlap in resource requirements can suffer from slowdowns as high as \textbf{1.5x}. On the other hand, scheduler that use feedback-driven approach first uses performance counters to measure IPC/Cache miss and {\it then} undertakes scheduling decision perform even worse, having performance degradation of more than \textbf{3.5x}. The reason for this that \textbf{Merlin} \cite{priyanka:socc14}  reacts to the resource requirements and suffers due to detection lag and other reasons mentioned earlier. Therefore, in order to maximize throughput, we need to act pro-actively, i.e anticipate the resource demand before it happens and act on its onset, which forms the primary objective of \textit{Beacons} Framework.

\subsection{Overall Framework}
\begin{figure*}
  \includegraphics[width=1.0\linewidth]{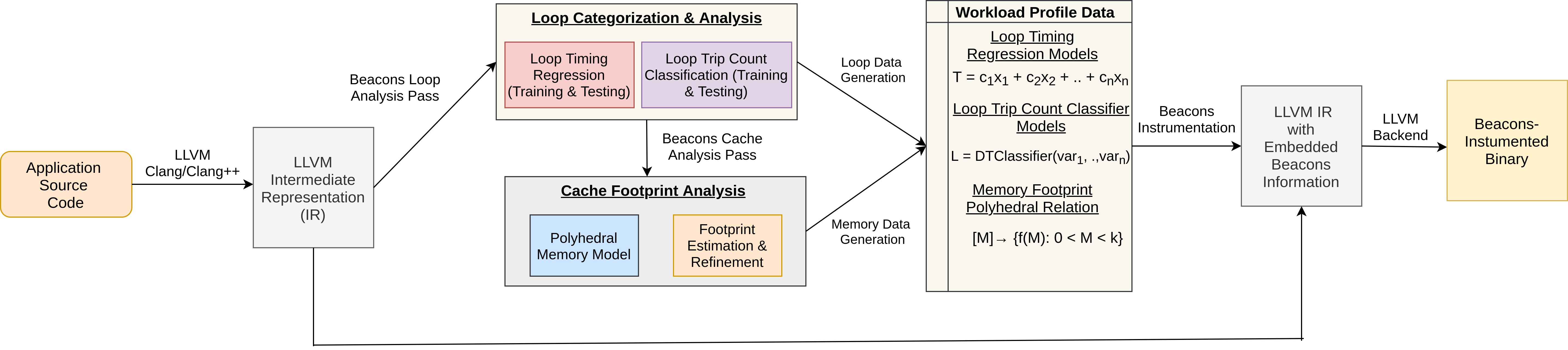}
  \caption{\small Beacons Compilation Component Workflow. The source code is translated into IR with LLVM front-end and is subjected to various `beacon' compiler passes to generate the loop and memory profile information. The IR is then instrumented with beacons information to broadcast information at the runtime to the Beacons Scheduler with the help of Beacons library. }\label{fig:system}
\end{figure*}

Beacons framework consists of a \textbf{Compilation Component} (Fig. \ref{fig:system}), which in turn comprises  of a sequence of compilation, training, and re-compilation steps to instrument the application; all of these steps are applied statically. Several challenges are overcome by to produce the above information at loop entrances. Beacons consist of statically inserted code at the entrances of the loops that evaluate closed-form formulae \& static learning models for predicting various artifacts related to loop's resource requirement. Other aspects, e.g. the amount of time a region executes, can be predicted \& enhanced by employing machine learning mechanisms that involves training. During runtime this information is evaluated by beacons using actual dynamic values and is conveyed to the beacons scheduler through library calls, which form the \textbf{Beacons Runtime Component} (Fig. \ref{fig:system}). The scheduler aggregates and analyzes this information across all the processes, constructing a global picture about contentions, the sensitivity to them, and moreover the duration of contentions. For the purposes of this paper, the resource under contention is the last level cache and memory bandwidth shared by all the processes in the machine. Since the resource demand and duration (loop timings) information is known to the scheduler {\it before} the loop executes, the scheduler is able to take pro-active scheduling decisions overcoming the afore-mentioned limitations of reactive approaches. 

 \begin{figure}[!ht]
 \centering\includegraphics[width=1.0\columnwidth]{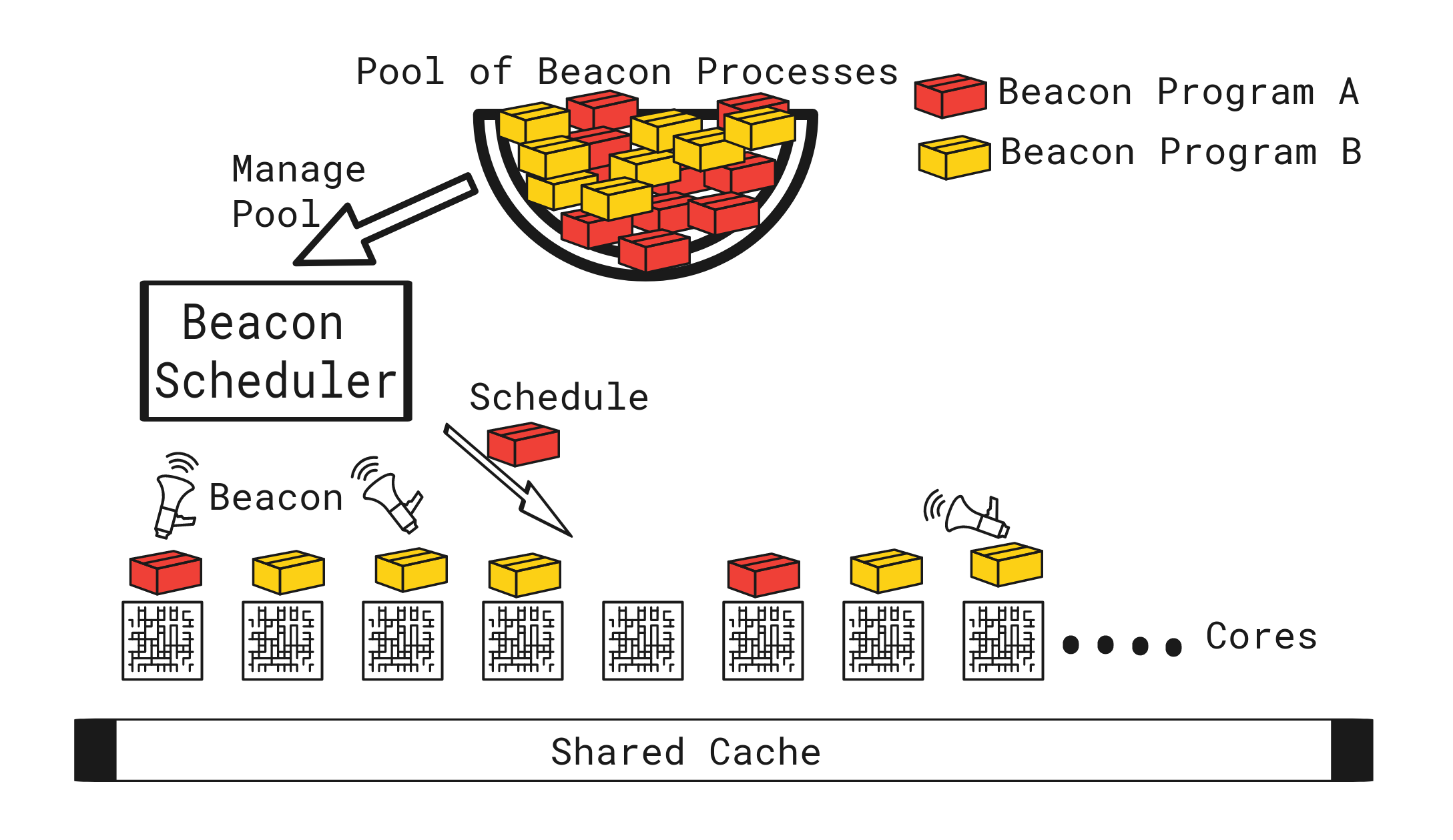}
 \caption{\small Beacons Runtime Component}
 \label{fig:bes_runtime}
 \end{figure}
\section{Beacons Compilation Component}
\label{beacons:compiler}
The Beacons Compilation Component (Fig. ~\ref{fig:system}) is responsible for instrumenting beacons in an application in order to guide a scheduler through upcoming regions. The regions in code are analyzed at the granularity of loop nests. During the compilation phase, the applications are first profiled to generate sample loop timings through a compiler pass called the \textit{beacons loop analysis pass}. This pass runs the program on a subset of application inputs, which form the training input set. These timings are then used for training a linear regression equation which establishes the loop timing as a function of the total iterations of the loop (\textbf{trip-count}). Such an approach directly handles simple loops whose trip-counts can be predicted statically from its loop bounds. However, to enable analysis of loops whose trip-counts are unknown at compile time, we develop a loop categorization framework that unifies both data and control flow of the loop bounds and its predicates variables. We then predict the trip counts for such loops by using decision tree classifier and rule-based mechanisms, which is futher integrated with the loop timings to enhance its accuracy. The timing regression equation is then embedded before the loop, along with two other pieces: the memory footprint equation and the classification of the loop as either ``reuse'' or ``streaming'', based on its memory reference patterns. The memory footprints are estimated by polyhedral compiler analysis \cite{grosser2011polly}, while the reuse classification is done with the help of static reuse distance (SRD) analysis. 

The beacons are inserted before the outermost loops and then hoisted inter-procedurally. To rectify potential scheduling errors arising from timing and memory footprint inaccuracies, the beacon-based scheduler is augmented with performance counters which are used in cases of loop bounds which can not be estimated even by machine learning; in addition a completion beacon is inserted at loop exit points to signal the end of the loop region. We now describe each of these above aspects in detail in the following subsections.
\subsection{Loop Timing Analysis \& Categorization}
\label{bec:timing}
In order to predict the timing of a loop (nest) present in the application, we establish it as a function of its total iterations (trip-count). Our goal is to obtain a linear relationship of form $y = ax_1 + bx_2 + .. + kx_n$, where $x_1, x_2, ..., x_k$ are trip-counts of individual loops in a loop nest and $y$ is the execution time of the loop nest. The core-idea here is that the constants $a,b,..,k$ can then be learnt using linear-regression and this closed-form equation can be directly utilized to generate the loop-timing dynamically during runtime.   
\subsubsection{General Loop Modelling} \label{bec:genloopmodel}For a given loop with various instructions, the loop time depends on the number of loop iterations, i.e. \textbf{loop time} ($T$) is directly proportional to \textbf{loop iterations} ($N$). Therefore, $T \propto N \Rightarrow T = \alpha * N$, where proportionality constant $\alpha$ which represents the average execution time of the loop body. In a nested loop, the execution duration depends on the number of iterations encountered at each nesting level. Consider a loop nest consisting of $n$ individual loops with total iterations $N_1, N_2,...,N_n$ respectively. We observe that the loop nest duration ($T$) is a function of these loop iterations, i.e $ T = f(N_1, N_2,...,N_n)$. As far as the loop body is concerned, each instruction in the loop nest contributes to the loop time by the factor of the number of times the instruction is executed. Thus, a loop can be analyzed by determining the nesting level of each of the basic blocks in its body and then multiplying the timing of that basic block with its execution frequency determined by the trip counts of surrounding loops. Thus, timing equation can be represented as: $T \propto g_1(N_1)+g_2(N_1*N_2)+...+ g_n(N_1*N_2*...*N_n)$, where $g_1(N_1)$ is the function of time contributed by outermost loop or the basic block belonging to the outermost loop with $N_1$ iterations, $g_2(N_1*N_2)$ is the function of time contributed by the second loop at the nesting level 2 and so on. Removing the proportionality sign, this equation can be rewritten as:

\begin{equation}\label{equation:lm}
    T = c_0 + c_1*N_1 + c_2*(N_1*N_2)+...+ c_n*(N_1*N_2...N_n)
\end{equation}
where $c_0$ is the constant term. Eq. \ref{equation:lm} is a linear equation in terms of each loop iteration $N_k$. Therefore, we use linear regression to learn the coefficients for each of the loop iterations in the loop nest to predict the loop timings.




\subsubsection{Estimating Precise Loop Iterations}
\begin{figure} [!ht]
\centering\includegraphics[width=1.0\columnwidth]{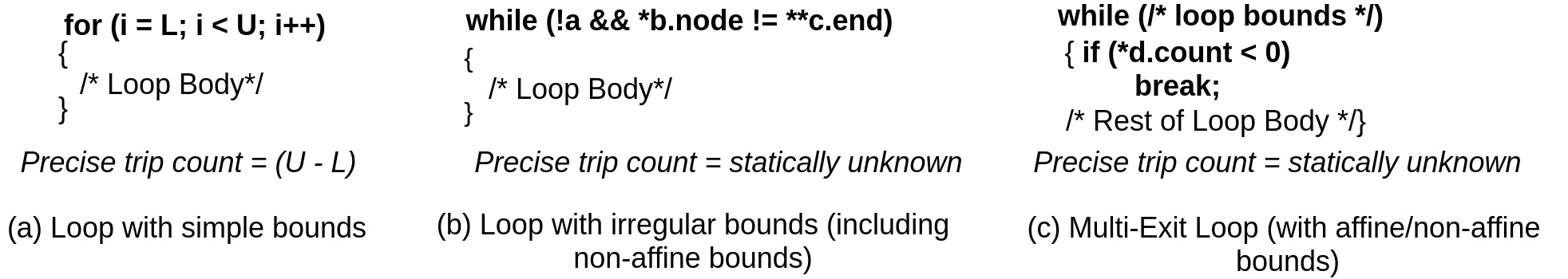}
\caption{Examples of different instances of loop for predicting trip counts. Each of these instances can be generalized to nested loops as well.}
\label{fig:loopex}
\end{figure}

:\label{bec:iter} The loop timing model described above depends on the compile time knowledge of the exact loop iterations, i.e the \textit{trip-count} associated with the loop. For certain kinds of regular loops, it's simple to statically determine the total number of iterations for loops with regular bounds (Fig. \ref{fig:loopex}{\color{red}a}). Such loops can be normalized using the loop-simplify pass in the LLVM \cite{Lattner:2004:LCF:977395.977673} compiler. Loop normalization converts a loop to the form with lower bound set to zero and step increment by one to reach an upper bound. The upper bound of the loop is now equal to the number of loop iterations and can be easily extracted from the loop and plugged into Eq. \ref{equation:lm} for timing prediction. In practice, however, loops need not adhere to this template. Real-world workloads have loops with irregular bounds (Fig. \ref{fig:loopex}{\color{red}b}) and non-uniform trip count that are often input-dependent, which makes it harder to analyze their trip count statically. This also extends to loop nests (either with affine or non-affine bounds), having control flow statements that break out iterations transferring the control out of the loop body (Fig. \ref{fig:loopex}{\color{red}c}). Therefore, to estimate precise loop iterations we need to develop a framework that captures both input data flow characteristics and control flow behaviour of the entire loop nest. 

\begin{figure} [!ht]
\centering\includegraphics[width=1.0\columnwidth]{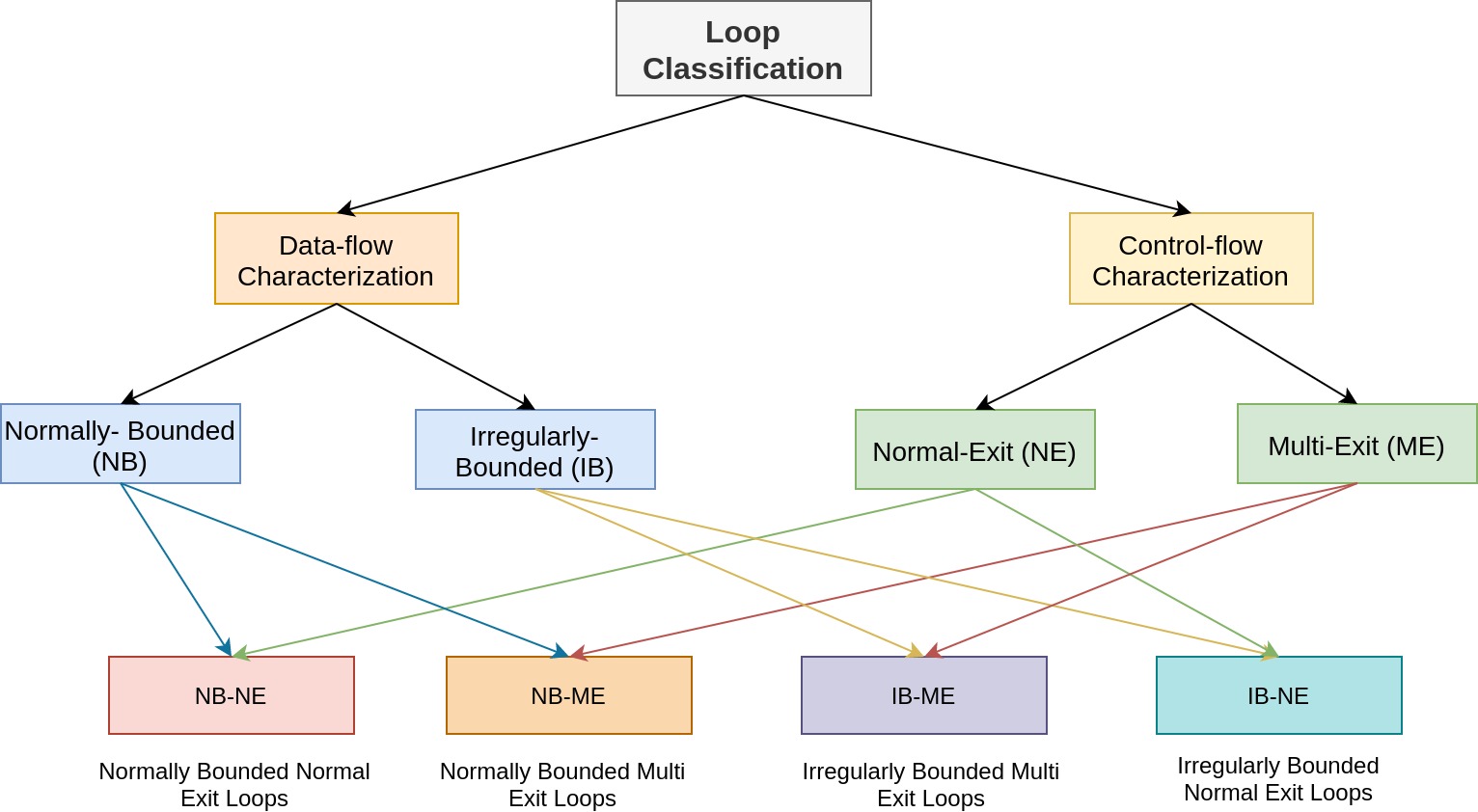}
\caption{Loop Classification Scheme with 4 classes of loop: NBNE, NBME, IBNE \& IBME}
\label{fig:looptax}
\end{figure}

\textbf{Loop Classification Scheme} In order to obtain trip counts for various loop-nests, we first categorize the loops by developing a classification scheme that takes into account both the data-flow \& control-flow characteristics of a loop. The data-flow characteristics of a loop can be determined by the loop bounds, while the control flow aspect can be captured by the nature of loop termination. Thus, each individual loops can either be \textit{Normal-bounded}/\textit{Irregularly-bounded} loops (data flow aspect) or \textit{Normal-Exit}/\textit{Multi-Exit} (control flow aspect) loops. Unifying the data-flow \& control-flow characterization results in a comprehensive framework that covers all possible loops types in real workloads (Fig. \ref{fig:looptax}) The resultant four classes of loops are \textit{Normally Bounded Normal Exit} (\textbf{NBNE}) loops, \textit{Normally Bounded Multi Exit} (\textbf{NBME}) loops, \textit{Irregularly Bounded Normal Exit} (\textbf{IBNE}) loops \& \textit{Irregularly Bounded Multi Exit} (\textbf{IBME}) loops. 

Based on this classification scheme, we develop a \textbf{Loop Categorization Algorithm} (Algo. \ref{algo:LCA}) that detects loops for each categories and invokes specialized trip count estimation mechanisms for each case. This algorithm first checks the bounds of the given loop. If the bound is entirely composed of numerical entities (integer, float, double, etc.), which allows a static estimation of symbolic upper bound the loop is classified as \textit{Normal-Bounded}. Alternatively, loops for which bounds contain non-numerical entities (pointer references, function calls, etc.) are considered as \textit{Irregularly-Bounded} Loops. To handle such cases, we propose \textbf{Upwards Exposed Control Backslicing} (UECB) Algorithm (Algo. \ref{algo:UECB}), which is a general framework that predicts trip counts based on the `upward-exposed' values of variables that decide the control flow in the loop, using learning mechanisms. It returns a learning model that estimates the loop trip counts based on a loop's critical variables.

\begin{algorithm}[htb]
\footnotesize
\SetAlgoLined
\textbf{Input:} Function F\\
\KwResult{Determine the precise trip count (in terms of expression or linear model) for all the loops}
\For{each loop $L \in$ Function $F$}
{
$regular,normal \leftarrow true$\\
$exit\_points,critical\_bounds,critical\_predicates \leftarrow \phi$\\
$bounds \leftarrow L.getBounds()$\\
\For{each argument $arg \in bounds$}
{ \If{type(arg) != num}
     {$regular \leftarrow false$\\
     $critical\_bounds.push\_back(arg)$}}
\For{each basicblock $bb \in L$}
{ \If{(bb.successor $\notin L$) and (bb != L.header) } 
     {$normal \leftarrow false$\\
     exit\_points.push\_back(bb)\\
     \For{each point $P \in exit\_points$}
     {\For{each variable $v \in P$}{critical\_predicates.push\_back(v)}}}
}

\If{regular and normal}
{$tripCount \leftarrow (bound.upper() - bound.lower())$}

\If{regular and not normal}
{$Prediction Model \leftarrow UECB (critical\_predicates)$\\}

\If{not regular and normal}
{$Prediction Model \leftarrow UECB (critical\_bounds)$\\}

\If{not regular and not normal}
{$Prediction Model \leftarrow UECB (critical\_bounds \cup critical\_predicates)$\\}

}
 \caption{Loop Categorization Algorithm}
 \label{algo:LCA}
\end{algorithm}

After the detection of expected trip counts based on loop bounds, the loop bodies are analyzed for control-flow statements. If there are no statements in the loop body that exit out of the loop for \textit{Normal-Bounded} case, the loop is categorized as \textit{Normal-Exit} and the expected trip count is equal to symbolic trip count. Next, if the loop is \textit{Irregularly-Bounded}, its expected bound is learnt using UECB algorithm described below. The cases of multiple exits in the body are classified as \textit{Multi-Exit} and the critical variables present in the predicates are extracted and passed on to the UECB algorithm to generate the respective model.

\textbf{UECB Algorithm}: The first step in UECB is to identify \textit{critical} variables that dictate a loop's trip count, i.e variable that decides whether the control flow will go to the loop body or loop exit. These variables can be either irregular loop bounds, or part of predicates that break out of loop. The core idea here is to use a learning model that can be trained on these variables to estimate the loop's trip count. The trained model can then be embedded in the beacon associated with the loop nest. However, since the inserted beacon exists at the entrance of loop nests, the trip count prediction must be done only with the variables \& their definitions that are live at that point. Such variables are called \textit{Out-of-loop} variables, i.e these variables are live at outermost loop header and their definitions that come from outside the loop body. Therefore, it is essential to back-propagate the \textit{critical} variables in order to express them in terms of \textit{Out-of-loop} variables. UECB achieves this back-propagation by following a stack-based approach that involves analyzing the \textit{critical} variables and the upwards-exposed definitions on which they depend in terms of their backward slice. When the program is run on representative inputs, these \textit{out-of-loop} variables are logged to generate the training data for the learning model. 

\begin{algorithm}
\footnotesize
\SetAlgoLined
\textbf{Input:} Set of Critical Variables C \\
\KwResult{Classifier Model M that estimates precise trip count based on unseen values of critical variables}
$model\_parameters, worklist = \{\}$\\
\For{each variable $v_c \in C$}
{$def\_set \leftarrow getallDefinitions(v_c)$\\
\For{each definition $d \in def\_set$}
{worklist.push\_back(d)}
}
\While{worklist is not empty}
{
  $d \leftarrow$ remove a definition from the worklist\\
  \For{each operand $op \in d$ }
  { \eIf{op is out-of-loop variable}
        {model\_parameters.push\_back(op)}
        {$def\_set \leftarrow getallDefinitions(op)$\\
         \For{each definition $d \in def\_set$}
           {worklist.push\_back(op)}
        }
  }
}
Generate Training \& Testing data for $model\_parameters$\\
\eIf{Total Data Points $>$ Threshold}
{Train Decision-Tree Classifier $M (model\_parameters)$}
{Obtain Rule-based Prediction Model $M (model\_parameters)$}
 \caption{Upwards Exposed Control Backslicing (UECB) Algorithm}
 \label{algo:UECB}
\end{algorithm}
The learning model used to predict loop trip counts based on \textit{critical} variables is the decision tree classifier. The input set for the entire program is divided into two parts for training \& testing respectively. This model is then embedded in the beacon associated with the outer-loop and is used to predict the precise loop trip count when it's invoked. 

\textbf{Loops not suitable for machine learning}: UECB Algorithm generates training data by logging the values of critical variables upon various loop invocations. However, if the loop is invoked only small number of times ($<10$), then it's impractical to train decision trees for predicting trip-counts of such loops. This is because machine learning models require the training and testing dataset to be sufficiently large in order to provide meaningful predictions. Thus, for predicting trip-counts of loops that are not invoked enough times to train a classifier model, we use \textbf{Rule-Based} Trip Count Prediction. The core idea here is that \textit{expected trip-count} is within the one standard deviation of the mean trip-count of all the rules. The rule-based mechanism is preferred over classifier model, when the number of data-points are lesser than hyper-parameter \textit{threshold} value ($\sim 5$).



After the expected trip count for various classes of loop are obtained, either by simple analysis (NBNE) or by classifier models (NBME/IBNE/IBME) from the Loop Categorization algorithm (Algo. \ref{algo:LCA}), they are integrated with timing equation (Eq. \ref{equation:lm}) to enhance the loop-timing predictions through normal regression models. 

\subsection{Memory Footprint Analysis}
\label{sec:memfp}
The footprint indicates the amount of cache that will be occupied by a loop. Memory footprint analysis consists of two parts - (a) calculating the memory footprint of the loop, and (b) classifying a loop as a \textit{reuse-oriented} loop or a \textit{streaming} (which exhibits little or no reuse) loop.


\subsubsection{Calculating Memory Footprint} For a given loop, its memory footprint is estimated  based on polyhedral analysis, which is a static program analysis performed on LLVM intermediate representation (IR). For each memory access statement in the loop, a polyhedral access relation is constructed to describe the accessed data points of the statement across loop iterations. An access relation describes a map from the loop iteration to the data point accessed in that iteration. It contains three pieces of information: 1) parameters, which are compile-time unknown constants, 2) a map from the iteration to array index(es); and 3) a Presburger formula~\cite{verdoolaege2016presburger} describing the conditions when memory access is performed. Generally, parameters contain  all loop-invariant variables that are involved in either array index(es) or the Presburger formula, and the Presburger formula contains loop conditions. We currently ignore if-conditions enclosing memory access statements; we thus get an upper bound in terms of estimation of the memory footprints. For illustration, list~\ref{code:mfa} shows a loop with three memory accesses, with two of them accessing the same array but different elements. A polyhedral access relation is built for each of them. The polyhedral access relation for $d[2*i]$ is: $[N] \rightarrow \{ [i] \rightarrow [2*i]: 0 <= i <= N \}$, where $[N]$ specifies the upper-bound of the normalized loop. It is a compile-time unknown loop invariant since its value is not updated  across loop iterations. $[i] \rightarrow [2*i]$ is a map from the loop iteration to the accessed data point (simply array indexes). $0 <= i <= N $ is the Presburger formula with constraints about when the access relation is valid.

\begin{lstlisting}[xleftmargin=0.05\textwidth,caption={Memory 
Footprint Estimation Example},label=code:mfa]
for ( int i = 0 ; i <= N; ++ i ) {
    ... = a[i+3];
    d[2*i] = ...;
    d[3*i] = ...;
}
\end{lstlisting}

Based on the polyhedral access relations constructed for every memory access in the loop, the whole memory footprint for the loop can be computed leveraging polyhedral arithmetic. It simply counts the number of data elements in each polyhedral access relation set, and then adds them together. Instead of a constant number, the result of polyhedral arithmetic is an expression of parameters. For $d[2*i]$, its counting expression generated using polyhedral arithmetic is: $[N] \rightarrow \{ (1 + N) : N >= 0 \}$. Therefore, as long as the value of $N$ is available, the memory footprints of the loop can be estimated by evaluating  the expressions. In our framework, the value of $N$ is given by the expected trip count (through one of the five cases: NBNE or by classifier models (NBME/IBNE/IBME) or rule-based system as described in the last section). For statements that access the same arrays, e.g. $d[2*i]$ and $d[3*i]$, a union operation will first be performed to calculate the actual number of accessed elements as a function of compile-time unknown loop iterations and instrumented in the program. It is evaluated at runtime to get the expected memory footprint.
\subsubsection{Classifying Reuse and Streaming Loops}\label{bec:reuse} 
\begin{figure}[!ht]
\centering\includegraphics[width=1.0\columnwidth]{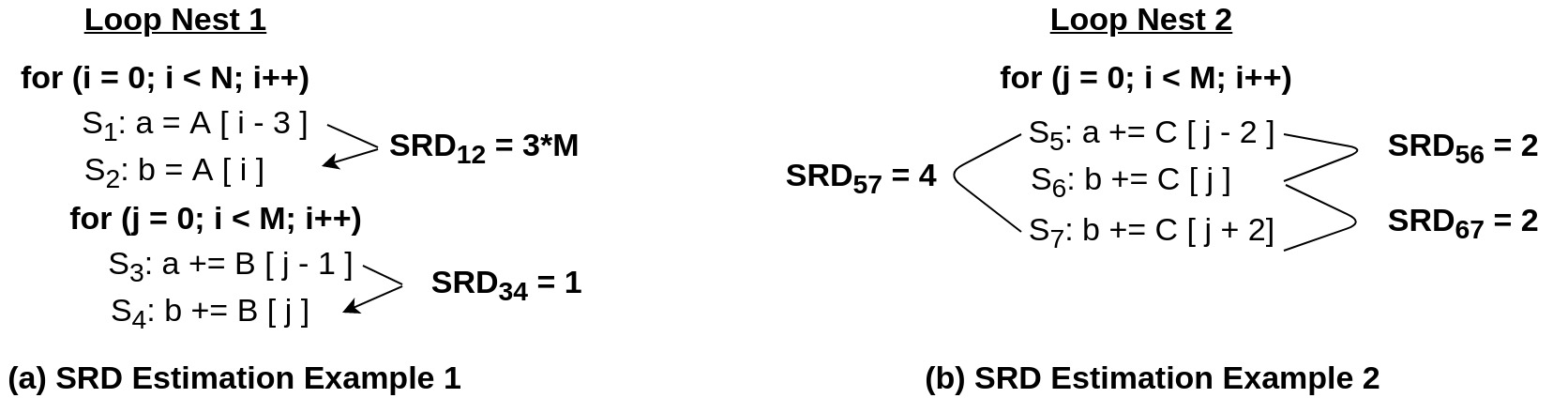}
\caption{\small SRD Estimation for two loop nests. Loop Nest 1 is classified as reuse, while Loop Nest 2 is streaming.}
\label{fig:loopSRD}
\end{figure}

A loop that reuses memory locations over a large number of iterations (large reuse distance) needs enough cache space to hold its working memory and might be sensitive to the misses caused, and a loop that streams data which is reused in next few iterations require almost no cache space at all and might be insensitive due to a small fixed reuse distance. For efficient utilization of cache, the scheduler must know this classification.  We classify the loops using \textbf{Static Reuse Distance (SRD)}, defined as the number of possible instructions between two accesses of a memory location. For example, in Fig. \ref{fig:loopSRD} the $SRD$ between statements ($S_1$, $S_2$) is in the order of $m*3$ because an access in instruction $S_1$ has to wait for $m$ instructions within the inner loop to cover the distance of three outer iterations between successive access of the same memory location. The $SRD$ between statements ($S_5$, $S_6$) is in the order of two, because the same memory is accessed after two iterations.

Any loops with a constant SRD, that is the distance between the accesses is covered within a few iterations of the same loop (e.g. the one between statements ($S_5$, $S_6$), ($S_5$, $S_7$) and ($S_6$, $S_7$) in Fig. \ref{fig:loopSRD}{\color{red}b}), can be classified as \textit{streaming}, because the memory locations must be in the cache for only a few (constant) iterations of the loop during which it is highly unlikely to be thrashed. More specifically, an SRD that involves an inner loop (e.g. the one between statements ($S_1$, $S_2$) in Fig. \ref{fig:loopSRD}{\color{red}a}) or
outer loop (e.g. between statements ($S_3$, $S_4$) in Fig. \ref{fig:loopSRD}{\color{red}a}) where a cache entry must wait in the cache over the duration of the entire loop that it is dependent on -- such loops are classified as \textit{reuse}. For example, array B must be in cache for the duration of the entire outer loop. Thus, we classify such loops in which the SRD is dependent on either an outer or inner loop as \textit{reuse loops} (reuse distance here is a function of normalized loop bound $N$, for example), and we classify the remaining loops (with small and constant reuse distance) as streaming loops. Indirect and irregular references such as a[b[i]] do not typically have a large reuse distance associated with them (compared to the sizes of modern caches) and it is impossible to analyze them at compile time; in our current approach, they are classified therefore as non-reuse references.

\subsection{Beacons Hoisting \& Insertion}
The beacon insertion compiler pass ensures that the beacons are hoisted at the entrances of outermost loops intra-procedurally. However, inter-procedural loops (function calls in loop body) can overload the scheduler with too many beacon calls. Hence, if the beacons are inside inter-procedural loops, then they are hoisted outside and also above the other call sites that are not inside loops along all paths. For hoisting the beacon call (\& embedding the inner-loop attributes), the inner loop bounds may not be available (or live) at the outermost points inter-procedurally. Thus, use expected loop bounds of the inner loops to calculate the beacon properties, timing information and memory footprint. Each interprocedural inner loop's memory footprint is added to the outermost inter-procedural loop's memory footprint. On the other hand, the loop timing is based only on the outermost inter-procedural loop. The beacon is classified as reuse if there is a single interprocedural nested loop. Unfortunately, such a conversion transforms many known/inferred beacons to unknown beacons. Also, hoisting is a repetitive process that stops once no beacons are inside inter-procedural loops. 

The decision trees are inserted as if-then-else blocks with the trip count value being passed to both the loop timing model and the memory footprint formula. The equations with the coefficients and loop bounds are instrumented at the preheader of the corresponding loop nests, followed by the memory footprint calculations. The variables that hold the timing and memory footprint values along with loop type (reuse or streaming) and beacon type are passed as arguments to the beacon library calls. Facilitated by the beacon library, the instrumented call fires a beacon with loop properties to the scheduler. We use shared memory for the beacon communications between the library and the scheduler. For every beacon, a loop completion beacon is inserted at either the exit points of the loop nest or after the call site for beacons hoisted above call site. The completion beacon sends no extra information other than signaling the completion of the loop phase and so that any sub-optimal scheduling decision can be rectified.
\section{Beacons Runtime Component}
\label{beacons_sch}

After the beacons and its attributes (loop timing regression models, trip count classifiers, memory footprint calculations and reuse classification) are instrumented in the application, it's the job of the Runtime Component to evaluate \& communicate the attributes during the execution. These evaluated attributes are communicated to the scheduler through a library interface that communicates with the scheduler. We refer to these function calls to the library as ``\textbf{beacon calls}''. A beacon call fired from an application writes the beacons attributes to a shared memory which is continuously polled by the scheduler. Beacon Scheduler analyzes this beacon information and acts proactively to the resource requirements among the co-executing processes, which sets it apart from state-of-art schedulers like CFS.  This establishes the communication between applications and scheduler (no special privileges required), and processes that can write to this shared memory are agreed upon during initialization with a key. The scheduler arbitrates the co-executing processes to maximize concurrency while simultaneously addressing the demand on the shared resources such as caches and memory bandwidth. Then goal of the scheduler is to facilitate efficient cache and memory bandwidth usage to improve system job throughput.

\textbf{Types of Beacon Calls}: Based on the system and application's execution state, there can be three distinct beacon library calls: \textit{Beacon\_Init} (to initialize the shared memory for writing the attributes), \textit{Beacon} (this writes the beacon attributes the shared memory) \& \textit{Loop\_Complete} (this signals the end of a loop for a process).  

\textbf{Beacons Call Classification}: Based on the precision of the attribute information, the beacon calls can either be classified into \textbf{Known/Inferred} Beacon Calls, where the loop trip-counts, timing and memory footprints are calculated via closed-form expressions with high accuracy, and \textbf{Unknown} Beacon Calls, where the attribute information is non-exact, expected values mainly calculated by rule-based trip-count estimation. This distinction helps us to identify potential impreciseness in application's resource requirement and allows us rectify certain sub-optimal scheduling decisions.  

\subsection{Beacons Scheduler (BES)}\label{bec:sch}
The beacon information sent by the applications is collected by the scheduler to devise a schedule with efficient resource usage. The beacon scheduler dynamically operates in two modes based on the loop data-reuse behaviour - reuse or streaming mode. The two modes corresponding to two types of reuse behaviour is to mitigate the adverse effects (resource demand overlaps) caused by multiple loops that co-execute together. Initially, the scheduler starts without a mode and schedules as many processes as required to fill up all the processors (cores) in the machine. One primary objective of the scheduler is to maximize system resource utilization and never keep any processors idle. 

The scheduler enters one of the two modes based on the first beacon it collects. Until the beacon call is fired, a process is treated as having no memory requirement and is referred to as a \textit{non-cache-pressure type process}. During the non-cache-pressure phase, the processes typically have memory footprint lower than the size of L1 private cache and do not affect the shared cache, unlike streaming or reuse type with cache requirements exceeding the L1 cache. Based on the timing information received from an incoming beacon, the scheduler estimates the time by which a certain loop should complete its execution. An important point to note here is although loop time values are obtained by compiling the process in isolation, the loop timing will still be similar even with multi-tenancy when scheduled by the beacon scheduler. This is because the scheduler ensures the avoidance of contention among the processes as detailed below.

\begin{figure} [!ht]
\centering\includegraphics[width=1.0\columnwidth]{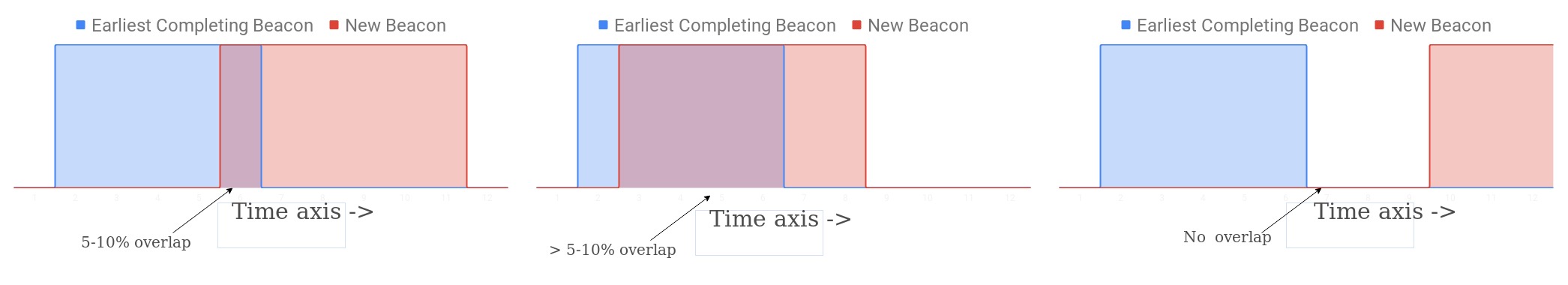}
\caption{Different timing Scenarios of Incoming Beacon}
\label{fig:scen}
\end{figure}

When any process fires a beacon, three possible timing scenarios (Fig~\ref{fig:scen}) can occur. In the first case, the completing (currently executing) beacon and the incoming beacon do not overlap (Fig~\ref{fig:scen} (right)). The completing beacon will relinquish its resources, which will be accordingly updated by the scheduler. The incoming beacon is then scheduled based on the resource availability.In the second case, the completing beacon and incoming beacon overlap for greater than 5-10\% (configurable) of the execution time (Fig~\ref{fig:scen} (middle)). Here, if the resources required by the incoming beacon is more than the available resources, then the incoming beacon process is descheduled and replaced with another process. Finally, in the third case where the overlap is less than 5-10\%, if the incoming beacons' resource requirement is satisfied on completion of the earliest beacon process, then the process is allowed to continue but with performance monitoring turned on. However, if the IPC of the beacon processes degrade, then the incoming beacon process is descheduled. Also, if the information is known to be imprecise (unknown beacons), then the scheduler turns on performance counters to rectify its actions. In each case the resource is either last level cache in reuse mode or memory bandwidth in stream mode. 

\begin{figure}
\centering\includegraphics[width=0.8\columnwidth]{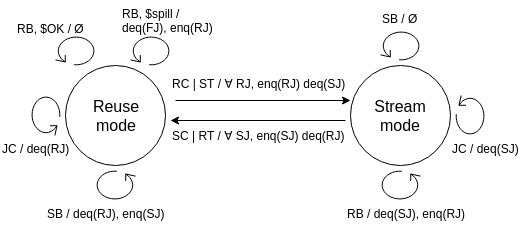}
\caption{A simplified Mealy state machine of the beacon scheduler. Key:
\textbf{B}eacon, \textbf{R}euse, \textbf{S}treaming, \textbf{F}iller,
\textbf{J}ob, \textbf{C}omplete, \textbf{T}hreshold,
\textbf{deq}ueue, \textbf{enq}ueue, \textbf{\$}(cache)}
\label{fig:fsm}
\end{figure}

\textbf{Reuse Mode}: The goal of the scheduler in reuse mode is to effectively utilize the cache by minimizing the execution overlap between the processes that are reuse bound. At any given time, the cores in the machine may be executing a mix of \textit{reuse loops} (\textbf{RJ}) that fit the cache and \textit{non-cache-pressure} (\textbf{FJ}) applications only as shown in the scheduler mealy machine Fig~\ref{fig:fsm}. If any of these FJ processes fires a \textit{reuse beacon} (\textbf{RB}), the scheduler first checks the memory information to ensure if the beacon fits in the available cache space or not. The scheduler only allows the process to continue if it fits in the available cache space. Once the reuse loop completes (loop\_completion beacon call), the process is re-classified as FJ again. If a FJ process fires a \textit{streaming beacon} (\textit{SB}), then the process is suspended and replaced by a suspended reuse process that fits in the cache. If no such suspended reuse process exists, then a non-cache-pressure process is scheduled. When all \textit{reuse loops are completed} (\textbf{RC}) or the number of suspended stream loops hits a \textit{threshold} (\textbf{ST}) (typically 90\% of the number of cores in the machine), then the remaining RJ processes are suspended, if any, and all the SJ processes are resumed and the scheduler switches to stream mode.

\textbf{Stream Mode}: A stream loop does not reuse its memory in the cache and hence is not disturbed by other co-executing processes as long as the system memory bandwidth is sufficient. The expected (mean) memory bandwidth ($\mu_{bw}$) of a stream loop can be calculated by using the memory footprint and the timing information as:  $\mu_{bw} = \frac{Memory Footprint}{Loop Time}$. In stream mode the scheduler schedules the SJ processes by replacing all other processes (RJ and FJ) as long as the Total Mean Memory Bandwidth (T$\mu_{bw}$) is less than the memory bandwidth of the machine. 

Any remaining core can only be occupied by a FJ process because a RJ process will get thrashed by the streaming applications. If a streaming loop completes, then it is replaced by a suspended streaming process when memory bandwidth is available. Otherwise, the process is allowed to continue as long as it does not fire a reuse beacon (\textbf{RB}). In other words, any non-streaming, non-cache-pressure application firing a reuse beacon is suspended and replaced by either a suspended streaming process or a non-cache-pressure application. When the number of such suspended reuse processes hits a threshold (RT), which is typically 10\% of the number of cores in the machine and based on whether the reuse processes can fill the cache, the scheduler switches from stream mode to reuse mode. An execution scenario is possible in which all streaming processes get suspended, all reuse processes are run, then after suspending more streaming jobs, all streaming processes are scheduled again in a batch, and so on.

\section{Evaluation}
\label{sec:eval}
Our goal is to evaluate Beacons Framework in environments, where throughput in terms of number of jobs completed per unit time or the time for completion of the batch is important and latency of each individual process itself is of not much value. A common example of such a scenario is a server conducting biological anomaly detection on thousands of patient data records (each patient as a new process) and with the goal to complete as many as possible to discover a particular pattern \cite{greensmith2010information}. 

\textbf{Experimental Platform}: The experiments were conducted on a Amazon Graviton2 (m6g.metal) machine, running Linux Ubuntu 20.4. The system has one socket with 64 processors and 32 MB L3 cache. We carried out our experiments on  60 processors, leaving the rest for other system applications to run smoothly and not interfere with our results. Beacons Compilation Component was implemented as unified set of compiler and profilling passes in LLVM 10.0. Machine learning library \textit{scikit-learn} was used to implement the classifier models.


\textbf{Benchmarks}: We evaluated our system on four sets of diverse workload suites, consisting of 53 individual benchmarks. We perform our experiments on: \textit{PolyBench} \cite{yuki2015polybench}, a numerical computational suite, consisting of linear algebra, matrix multiplication kernels, \textit{Rodinia} \cite{che2009rodinia}, which consists of graph traversal, dynamic programming applications from diverse domains like medical imaging, etc and on various popular machine learning benchmarks like \textit{AlexNet} ~\cite{hinton2012improving}, \textit{DenseNet201} \cite{huang2017densely}, \textit{Resnet101} \cite{he2016deep}, \textit{Resnet-18}, \textit{Resnet-152}, and \textit{VGG16} \cite{simonyan2014very} by training these networks on a subset of CIFAR-10 Dataset \cite{krizhevsky2009learning}. In addition to that, we perform experiments on well-known pre-trained networks like \textit{TinyNet}, \textit{Darknet} and \textit{RNN} (with 3 recurrent modules) for predicting data samples from CIFAR-10, Imagenet ~\cite{deng2009imagenet} and NLTK \cite{bird2009natural} corpus respectively.

\textbf{Methodology}: We run experiments and report all the benchmarks in Polybench. However, since the L1 data cache size is 32KB, the beacon calls are fired only if the loop memory footprint is above 32KB and if the loop predicted time is above 10ms. This is neccessary because the average processing time of loop complete, reuse, and stream beacons are 116us, 427us, and 292us respectively, and thus the loop timings below 10ms would be unneccessary for scheduling. \textit{Leukocyte} in \textit{Rodinia} is one such benchmark with all beacons statically removed because the expected memory footprint is lower than 32KB and hence we do not report the values here. Our ML workloads were created using \textit{Darknet} \cite{darknet13} and they were divided into training benchmarks and prediction benchmarks. The training benchmarks are well-known \textit{convolutional neural networks} (CNNs) and we ran them by training on a subset of Cifar-10 images. The prediction benchmarks are pre-trained models that were used to classify images from \textit{Imagenet} dataset. For all the benchmarks, input sets were divided into training \& test sets to obtain the loop-timing (regression) \& trip-count (classification) models. Table \ref{tab:ben} summarizes all the benchmarks used in our experiments. All the benchmarks were compiled with their defaults optimization flags (-O2/O3, -pthread, etc).
\begin{table}[htb]
\resizebox{\columnwidth}{!}{
\begin{tabular}{|l|l|l|}
\hline
\textbf{Benchmark Suite} & \textbf{Benchmarks} & \textbf{Dataset} \\ \hline
Polybench & \begin{tabular}[c]{@{}l@{}}2mm, 3mm, atax, bicg, mvt, gemm,gesummv,\\ symm, syr2k, syrk, trmm, cholesky, lu, \\ ludcmp, trisolv, correlation,covariance, \\ floyd-warshall, nussinov, deriche, adi, fdtd-2d, \\ heat-3d, jacobi-1d, seidel-2d\end{tabular} & \begin{tabular}[c]{@{}l@{}}SMALL\\ STANDARD\\ EXTRALARGE \\ (Training)\\ \\ LARGE (Testing)\end{tabular} \\ \hline
Rodinia & \begin{tabular}[c]{@{}l@{}}backprop, bfs, cfd, heartwall, hotspot, \\ hotspot3D, kmeans-serial, lavaMD, nn, \\ particlefilter,  srad\_v2\end{tabular} & \begin{tabular}[c]{@{}l@{}}Custom Inputs (Training \\ \& Testing)\end{tabular} \\ \hline
\begin{tabular}[c]{@{}l@{}}Convolutional Neural Network\\ (Training)\end{tabular} & \begin{tabular}[c]{@{}l@{}}Alexnet, Resnet-18, Resnet-101, Resnet-151,\\ VGG-16, Densenet-201\end{tabular} & \begin{tabular}[c]{@{}l@{}}CIFAR-10 (Training \\ \& Testing)\end{tabular} \\ \hline
\begin{tabular}[c]{@{}l@{}}Convolutional Neural Network\\ (Pre-Trained)\end{tabular} & TinyNet, Darknet3 & \begin{tabular}[c]{@{}l@{}}Imagenet (Training  \\ \& Testing)\end{tabular} \\ \hline
\end{tabular}}
\caption{Summary of Benchmarks used in our experiments}
\label{tab:ben}
\end{table}
\textbf{Designing Scheduling Jobs}:
Our mixes consists of large and small processes. We set a reasonable number of large processes (20-200) so the mix does not run for unusually large duration (hours) for experimentation although our scheduler is robust works long running processes. Once the large processes are executed, we inject 4-5 small processes per large process to simulate a real-life scenario of other processes getting added to mixes and trying to hog the cache. Thus, the scheduler can end up executing more than 10000 processes during the entire mix. In addition, our mixes are homogeneous so all processes are the same, but with different inputs only. We also created heterogeneous mixes of different applications but a homogeneous mix tends to be the worst case because all processes have the same phases and cache requirements and execute in somewhat of a phase synchrony causing each process to demand the same resources as other processes at the same time during its execution. 

\textbf{Baselines}: To evaluate the effectiveness of Beacons Scheduling mechanism, we compare it against - \textbf{(a)} Linux's CFS \cite{kobus2009completely} baseline, which is the widely-used standard scheduler in most computing environments, and \textbf{(b)} Merlin~\cite{priyanka:socc14}, which is a performance-counter based reactive scheduler (RES). The purpose of this comparison to investigate the necessity of compiler-guided scheduling vs traditional perf-based scheduling. Merlin first uses the cache misses per thousand instructions (MPKI) in LLC to determine memory intensity; then it estimates cache reuse by calculating the memory factor (MF), which is the ratio LLC/(LLC-1) MPKI. We use the same MF threshold as Merlin (0.6) to classify reuse and stream phases. 


\begin{figure} [!ht]
\centering\includegraphics[width=1.0\columnwidth]{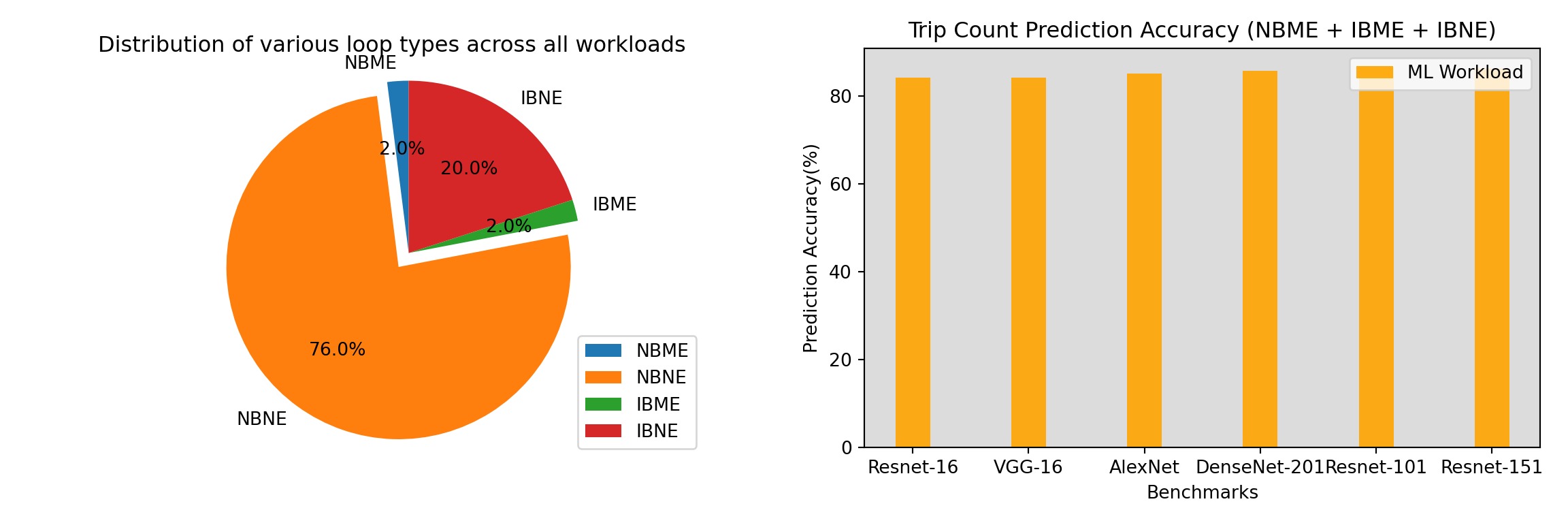}
\caption{Loop Distribution across Rodinia \& ML Workloads and their trip-count prediction accuracy.}
\label{fig:loopdistype}
\end{figure}

\subsection{Beacons Prediction Accuracies}
\textbf{Loop Trip Count Prediction Accuracy:}  Fig. \ref{fig:loopdistype} shows the distribution of different loop-classes for \textit{Rodinia} and \textit{Machine Learning} workloads. \textit{Polybench} is not shown in the pie-chart since it contains 100\% NBNE-type loops. \textit{Rodinia}'s loops are rule-based so there are no classifiers needed. The right part of Fig.\ref{fig:loopdistype} shows the accuracy of classifier-based models in predicting the trip counts. The average accuracy is \textbf{85.3\%}. The reason UECB achieves consistently high accuracy is because of the nature of these benchmarks. Machine Learning workloads can be broadly dissected into several constituent layers - convolutional layer, fully-connected layer, softmax-layer, dropout-layer, etc. Each of these layers perform an unique set of operations, that can be captured very well by classifier models employed by UECB. These models find the correlation between specific loop properties such as the trip counts, critical variable values and hence are able to perform well. Overall, trip counts for \textbf{60\%} of the loops in the workloads are predicted by \textit{classifier} models, while the rest are obtained by \textit{rule-based} models.

\begin{figure}
  \includegraphics[width=1.0\linewidth]{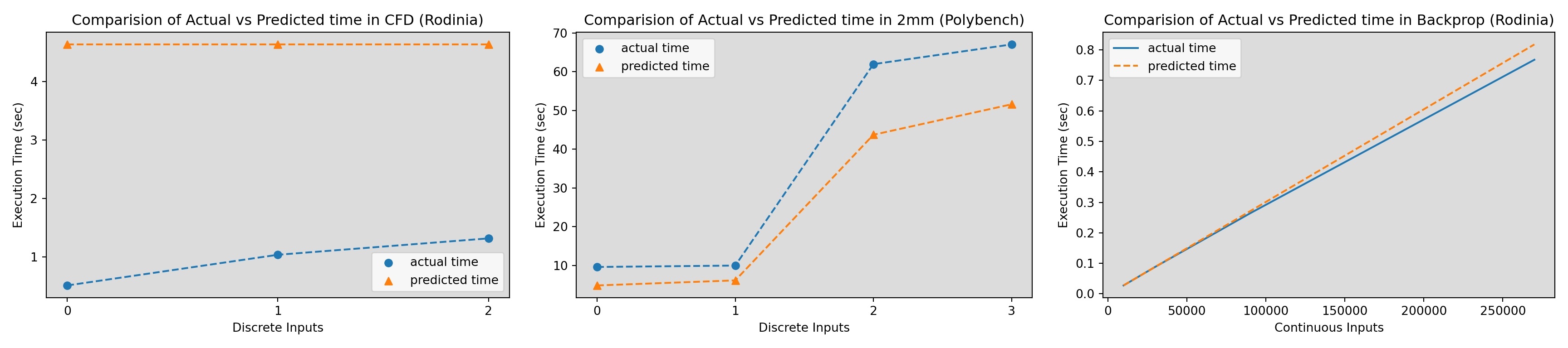}
  \caption{\small Three different cases of Loop Timing Accuracy: positive deviation (CFD), negative (2mm) and perfect fit (Backprop) }\label{fig:actvpred}
\end{figure}

\begin{figure*}
  \includegraphics[width=1.0\linewidth]{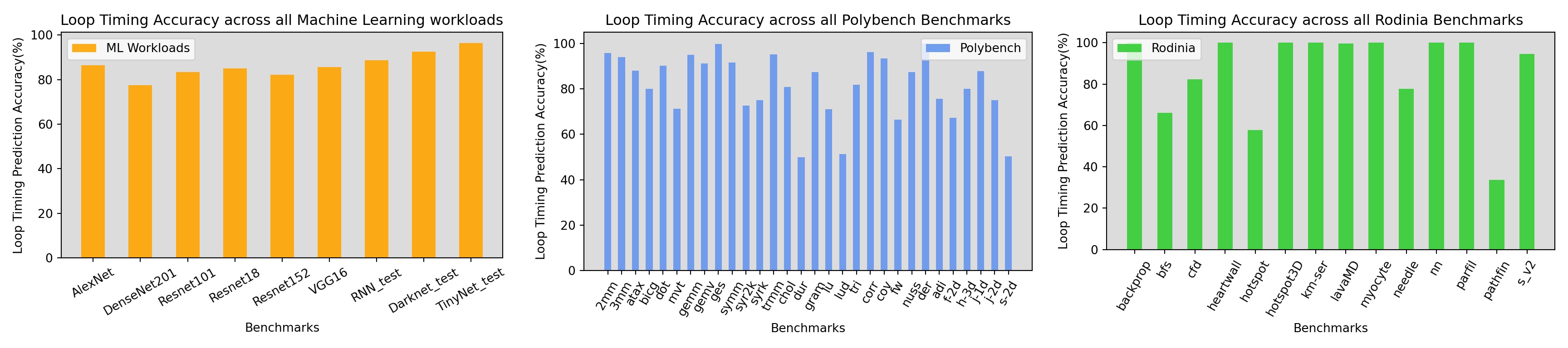}
  \caption{\small Loop Timing Accuracy across all three workloads suites}
  \label{fig:time1}
\end{figure*}

\textbf{Loop Timing Accuracy}: The overall loop timing accuracy for both \textit{unknown} and \textit{known} beacons collectively was \textbf{83\%} across all the workloads (Fig.\ref{fig:time1}). This accuracy can vary depending on different classes of loops and their respective training inputs for closed-form regression equation (Eq.\ref{equation:lm}). For known/inferred beacons (NBNE and other loops covered by UECB algorithm), the accuracy of timing info mainly depends on training. The regression fit in Eq.~\ref{equation:lm} works particularly well for continuous input sets, as the inputs can be monotonously increased or decreased to obtain a proper domain for the regression function. For example, in Fig.\ref{fig:actvpred}, \textit{backprop} takes in continuous integer inputs and the predicted curve matches closely with different testing inputs (mean-squared error $\mu = 0.057$). However, for applications with discrete inputs, the regression training becomes dependent on how representative the provided training inputs are of the actual application behaviour. Some benchmarks like \textit{hotspot} in Rodinia have five inputs (four used for training) that capture the behaviour of the loop. The precise loop curve overlaps almost exactly with the actual time curve, similar to Fig.\ref{fig:actvpred}. In contrast, a few cases had training inputs that are not sufficient enough to learn precise regression coefficients, thus decreasing the accuracy. For example, in 2mm, shown in Figure~\ref{fig:actvpred}, the predicted curve deviates for the fourth input (which is the test input). 

For some NBME, IBME and IBNE loops, the unknown beacons predicts a loop time that is loop bound obtained during regression runs. However, the actual loop time can deviate from the predicted time, resulting in either positive or negative error rate. These two cases are illustrated in Fig~\ref{fig:actvpred}. In \textit{CFD}, the timing information is generated by hoisting the expected values of inner loop bounds to a point above their inter-procedural outer loop in the beacon. As a result, we end up with a mean-squared error $\mu = 11.023$, which shows that ``unknown'' nature of these loops can sometimes cause unreliable prediction. However, to ensure that such cases do not impact the scheduling decisions, an end-of-loop beacon typically fires at the loop exit, which helps the scheduler correct its course. Ultimately, these few cases of expected predictions (with low accuracy) are still manageable mainly due to loop completion beacons.


\subsection{Throughput Improvement \& Analysis}
\begin{figure*}
  \includegraphics[width=1.0\linewidth]{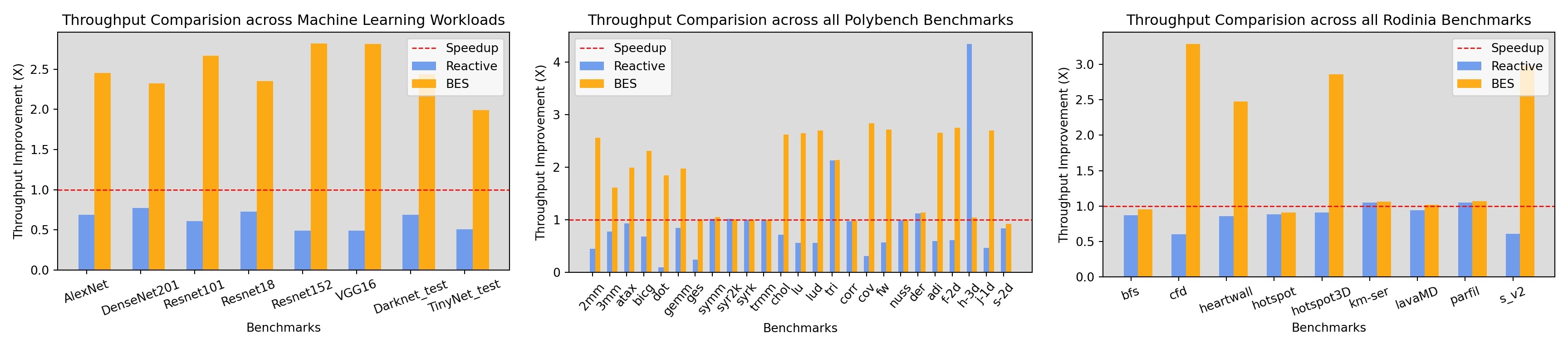}
  \caption{\small Throughput Improvement across benchmarks in Polybench, Rodinia \& Machine Learning Workloads. Both scheduler’s throughputs are normalized against the baseline Linux’s CFS Scheduler}\label{fig:thr}
\end{figure*}

The throughput of the system is calculated as the total time required by the scheduler to finish a fixed number of incoming jobs which is same as the average number of jobs completed in unit time when normalized with a baseline (CFS). The throughput of both the beacon scheduler and Merlin-based reactive scheduler normalized against CFS is shown in Fig. \ref{fig:thr}. Based on geometric mean average, we achieved speedup of \textbf{76.78\%} on Gravitron2 compared to the 33\% slowdown by the Merlin-based reactive scheduler. Among 45 evaluated benchmarks from Polybench, Rodinia, and modern ML workloads, we had significant throughput improvement for 28 of them with the modern ML workloads showing the most throughput improvement (\textbf{2.47x}) on geometric mean. Polybench and Rodinia have a modest improvement of 69.01\% and 51.46\% respectively. These improvements are for the worst-case mix of homogeneous processes that have the same cache phases.

The performance benefit is mainly attributed to the smart processes scheduling leveraging the knowledge of process' loop type at a given time. The throughput improvement can be divided into several categories based on its throughput improvement and are analyzed below: 
\begin{itemize}
    \item \textbf{Insignificant Improvement}: In four workloads, beacons performed worse or got similar timings compared against CFS baseline. These workloads were typically dominated by streaming loops ($>95\%$), and thus majority of execution time was spend on streaming loops. In this scenario, beacons' scheduling is similar to CFS, where the resource demand are low and uniform. Adding the small scheduling and beacons call overheads, we end up with timing similar to CFS' or small slowdowns ($\leq9\%$).
    \item \textbf{Small Improvement} (10\% - 50\%): The majority of applications within this category tend to have small amount of reuse-based loops ($~5\% - 10\%$). Most of the reuse loops typically have smaller footprints or short duration. Deriche is a special case where reuse loops occur alternatively between streaming loops. Therefore, the minimal performance improvement is obtained from the data-reuse in alternate loops, but the frequent streaming loops prevents the inter-loop data reuse. While CFS can save on memory accesses across loops (because CFS does not preempt the application), BES gains performance during the reuse loops, leading to smaller overall improvement.
    \item \textbf{Medium Improvement} (50\% - 2x): Applications within this category spend a considerable amount of time within reuse loops. Bicg, gemm and atax have a majority of streaming loops but the reuse loop executes for the longest duration. On the other hand, trisolv has more reuse loops (but lesser duration) which allows beacons scheduler to be in reuse mode for a longer period of time collectively, and benefits from the proactive scheduling.
    \item \textbf{High Improvement} (2x - 3x): Applications run for longer durations and spend a majority of their time within reuse loops (although they have a mix of streaming and reuse loops in some cases). Evidently, it it's no surprise that all the ML workloads fall in this category. These workloads are mostly training or predicting so they have to continuously reuse their neuron weights during the execution. Aside from the neuron weights, they typically use matrix computations, which are further reused to do back-propagation. Both the matrices and weights account for a considerable amount of data, as the large networks like Alexnet has eight layers (especially five convolutional layers). Thus, beacons schedules the processes to enable fast matrix multiplications due to less trashing of the cache.
\end{itemize}

\begin{figure} [!ht]
\centering\includegraphics[width=1.0\columnwidth]{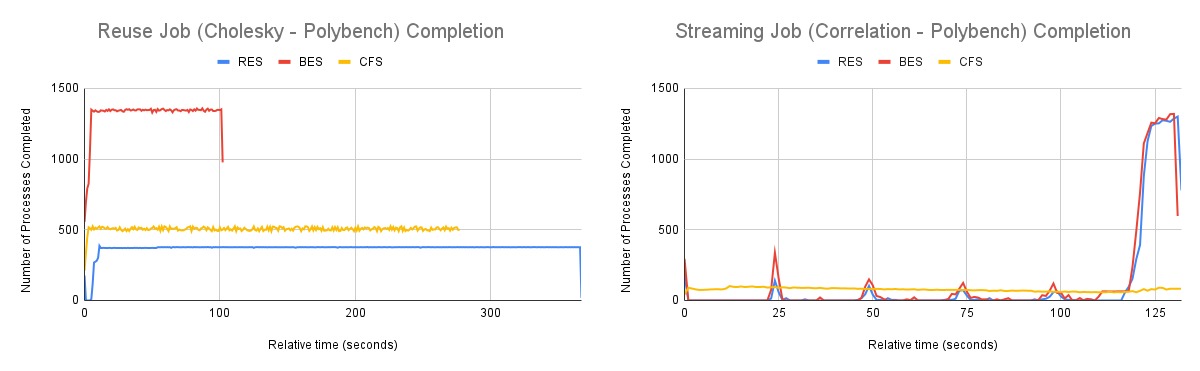}
\caption{Histograms for the job completion times of CFS and the beacon-enabled scheduler (BES) for cholesky (left) and correlation (right). The X axis represents discrete timesteps, and the Y axis is a count of the number of jobs that completed within a given timestep}
\label{fig:hist}
\end{figure}

We finally present three interesting job completion timelines of \textit{Cholesky} (which showed substantial benefits with beacons), juxtaposed with those of \textit{Correlation} (which had no noticeable benefit)(Fig. \ref{fig:hist}) with three scheduler - \textbf{CFS}, Beacons' scheduler (\textbf{BES}) and Merlin-based reactive scheduler (\textbf{RES}). In \textit{Cholesky}, BES starts with the same jobs as CFS but soon replaces some of the reuse jobs with other non-cache-pressure types to avoid cache overflow, unlike CFS which takes longer to retire its first jobs. BES later on intelligently places reuse and non-cache-pressure jobs to maintain high throughput, whereas CFS keeps scheduling both the non-cache-pressure types and cache-pressure types throughout the execution. Compared to these two, RES keeps on moving both cache-pressure types and non-cache-pressure types rather than prioritizing a particular type of them at the beginning.  In the case of correlation, the jobs are within the cache size limit and this can be seen by the two graphs (BES and RES) being interleaved. Thus, BES does not do anything differently from RES, and both complete their workloads at roughly the same time. BES and RES try to prioritize one group and this leads to no completions then a spike of processes completing especially in the tail end of the mix that saw ~1200 processes completing every second. On the other hand, just like in cholesky, CFS has no knowledge of the types and does not do it in a batch so CFS keep on scheduling and preempting processes to allow a constant amount of ~80 processes to complete at each second. Thus, its line is mostly horizontal. In summary, in \textit{Cholesky}, by limiting cache pressure through careful scheduling, BES wins over others whereas in \textit{Correlation}, it performs no worse.

\section{Related Works}
\label{sec:related}

\textbf{Conservative scheduling}~\cite{yang2003conservative} presents a learning-based technique for load prediction on a processing node. The load information at a processing node over time is extrapolated to predict load at a future time. Task scheduling is done based on predicted future load over a time window. A compiler-driven approach to reduce power consumption by inserting statements to shut down functional units through a profile-driven approach in areas of a program where no access to the units happens is proposed in \cite{rele2002optimizing}; it uses a profile-driven approach that predicts the future execution paths and the latency.

\textbf{Profile-based scheduling}: Prediction of applications' upcoming phases by using a combination of offline and online profiling is proposed in ~\cite{shen2004locality,padmanabha2013trace}. Similarly, another approach~\cite{sashaThroughputScheduling} uses a reuse distance model for simple caches calculated by profiling on a simulator for predicting L2 cache usage. A cache-aware scheduling algorithm is proposed in \cite{DBLP:conf/usenix/FedorovaSSN05}. It monitors cache usage of threads at runtime and attempts to schedule groups of threads that stay within the cache threshold. It is not compiler-driven, nor sensitive to phase changes within the threads. Several efforts have focused on the development of scheduling infrastructure for the shared server platforms~\cite{bubbleflux,bubbleup,deepdive,fedorova,priyanka:socc14}. A key feature of these efforts is their use of observation-based methods (i.e. reactive approaches) to establish resource contention (e.g. for caches, memory bandwidth, or other platform resources) and to further determine the interference at runtime, and/or to assess the workloads' sensitivity to the contended resource(s) by profiling.
\section{Conclusion}
\label{sec:conc}

In this work, we propose a compiler-directed approach for proactively carrying out 
scheduling. The key insight is that the compiler
produces predictions in terms of loop timings and underlying
memory footprints along with the type of loop: reuse oriented
vs streaming which are used to make scheduling decisions. A new framework based on the combination of static analysis coupled with ML models was developed to deal with irregular loops with statically non-determinable trip counts and with multiple loop exits. It was shown that this framework is able to successfully predict such loops with \textbf{85\%} accuracy. A prototype implementation of the framework demonstrates significant
improvements in throughtput over CFS by an average of \textbf{76.78\%} with up to \textbf{3.25x} on Graviton2 for consolidated workloads. The value of prediction
was also demonstrated over a reactive framework which under-performed CFS by \textbf{9\%}. Thus, to conclude, predictions help pro-activeness in terms of scheduling decisions which lead to significant improvements in throughput for a variety of workloads. 

\bibliographystyle{plain}
\bibliography{references}

\end{document}